\documentclass[runningheads]{svmult}

\usepackage{makeidx}   
\usepackage{graphicx}  
\usepackage{subeqnar}  
\usepackage{multicol}  
\usepackage{physprbb}  
\makeindex             

\def\etal{{\it et al.}}
\def\ltsima{$\; \buildrel < \over \sim \;$}
\def\simlt{\lower.5ex\hbox{\ltsima}}
\def\gtsima{$\; \buildrel > \over \sim \;$}
\def\simgt{\lower.5ex\hbox{\gtsima}}

\begin{document}

\title*{Optical Spectra and Light Curves of Supernovae}

\toctitle{Optical Spectra and Light Curves
\protect\newline of Supernovae}

\titlerunning{Optical Observations of Supernovae}

\author{Alexei V. Filippenko}
\authorrunning{Alexei V. Filippenko}
%
%
\institute{Department of Astronomy, University of California, Berkeley,
 CA 94720-3411 USA\\ e-mail: alex@astro.berkeley.edu
}

\maketitle              

\begin{abstract}

I review recent optical observations of supernovae (SNe) conducted by my
group. The Lick Observatory Supernova Search with the 0.76-m Katzman Automatic
Imaging Telescope is currently the world's most successful search for nearby
SNe. We also use this telescope to obtain multicolor light curves of SNe. One
of the more interesting SNe we discovered is SN 2000cx, which differs from all
previously observed SNe~Ia. Another very strange SN~Ia that we studied is SN
2002cx, many of whose properties are opposite those of SN 2000cx. Extensive
data on SNe~II-P 1999em and 1999gi were used to derive distances with the
expanding photosphere method. Results from spectropolarimetry suggest that the
deeper we peer into the ejecta of core-collapse SNe, the greater the
asphericity. We are using {\it Hubble Space Telescope} data to identify, or set
limits on, the progenitors of core-collapse SNe.

\end{abstract}

\section{The Lick Observatory Supernova Search (LOSS)}

   In 1989, my team began to work on developing a robotic telescope for CCD
imaging of relatively faint objects. The history of the project is discussed in
several papers (e.g., Filippenko \etal\ 2001; Richmond, Treffers, \& Filippenko
1993), and several prototypes were used over the years. In 1996, we achieved
first light with our present instrument, the 0.76-m Katzman Automatic Imaging
Telescope (KAIT) at Lick Observatory on Mt. Hamilton, California. It took the
better part of another year to eliminate most of the remaining bugs in the
system, and useful scientific results started appearing in 1997. Absolutely
vital contributions to the programming and to the observing strategy were made
by Dr. Weidong Li, who joined my group in 1997.

 KAIT is a fully robotic instrument whose control system checks the weather,
opens the dome, points to the desired objects, acquires guide stars (in the
case of long exposures), exposes, stores the data, and manipulates the data
automatically, all without human intervention. We reach a limit of $\sim19$ mag
($4\sigma$) in 25-s unfiltered, unguided exposures, while 5-min guided
exposures yield $R \approx 20$ mag.  KAIT acquires well-sampled, long-term
light curves of SNe and other variable or ephemeral objects --- projects that
are difficult to conduct at other observatories having a large number of users
with different interests.

   One of our main goals is to discover nearby SNe to be
used for a variety of studies.  Special emphasis is placed on finding them well
before maximum brightness. Although the original sample of our Lick Observatory
Supernova Search (LOSS; Li \etal\ 2000; Filippenko \etal\ 2001) had only about
5000 galaxies, in the year 2000 we increased the sample to $\sim 14,000$
galaxies (most with redshift $\simlt 10,000$ km s$^{-1}$), separated into three
subsets (observing baselines of 2 days for about 100 galaxies, 3--6 days for
$\sim 3000$ galaxies, and 7--14 days for $\sim 11,000$ galaxies). We are able
to observe $\sim 1000$ galaxies per night in unfiltered mode. Our software
automatically subtracts new images from old ones (after registering, scaling to
account for clouds, convolving to match the point-spread-functions, etc.), and
identifies SN candidates (Fig. 1) which are subsequently examined and reported
to the Central Bureau for Astronomical Telegrams by numerous undergraduate
research assistants in my group, working with Weidong.

\begin{center}
\scalebox{0.22}{\includegraphics{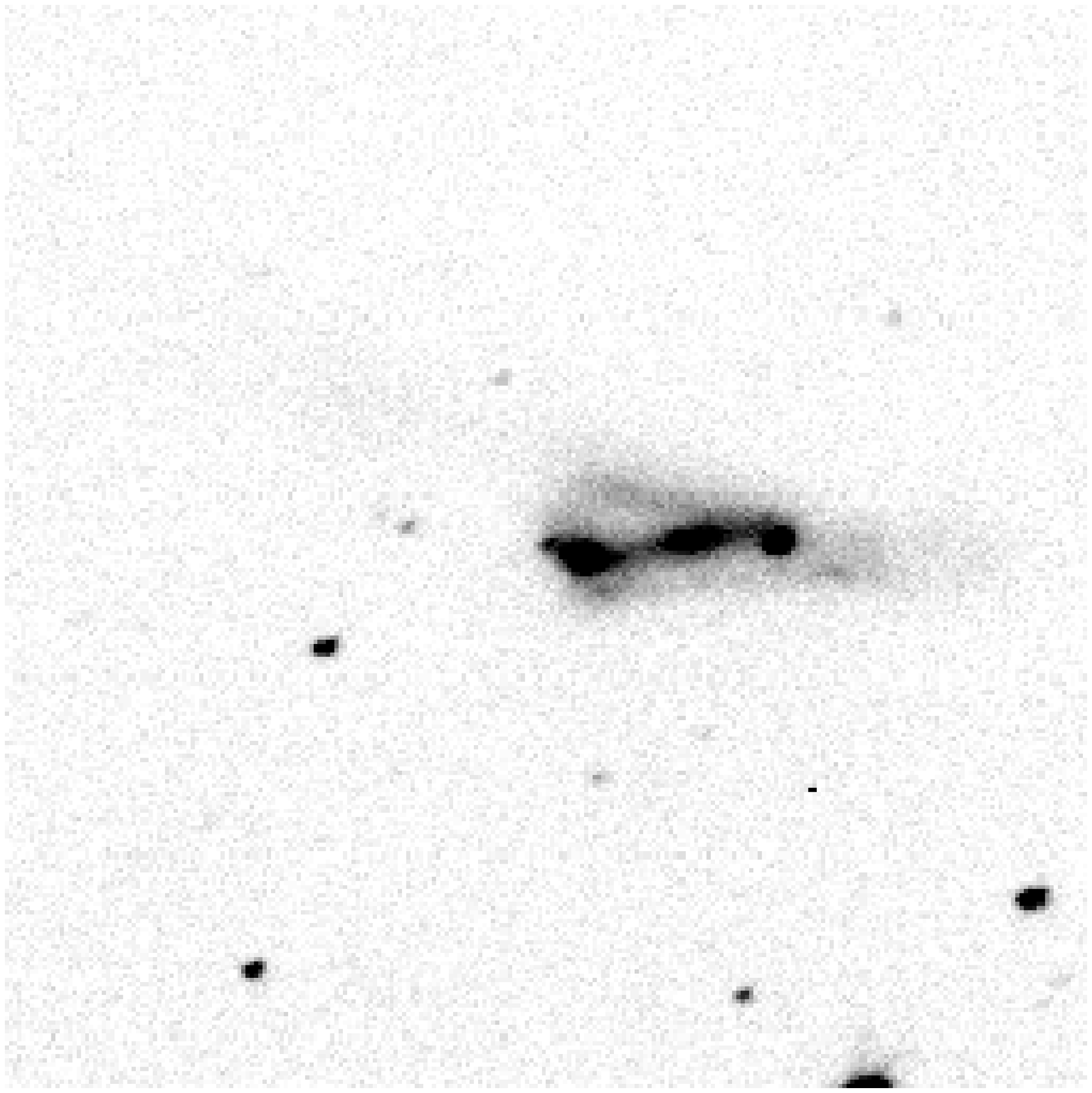}}
\scalebox{0.22}{\includegraphics{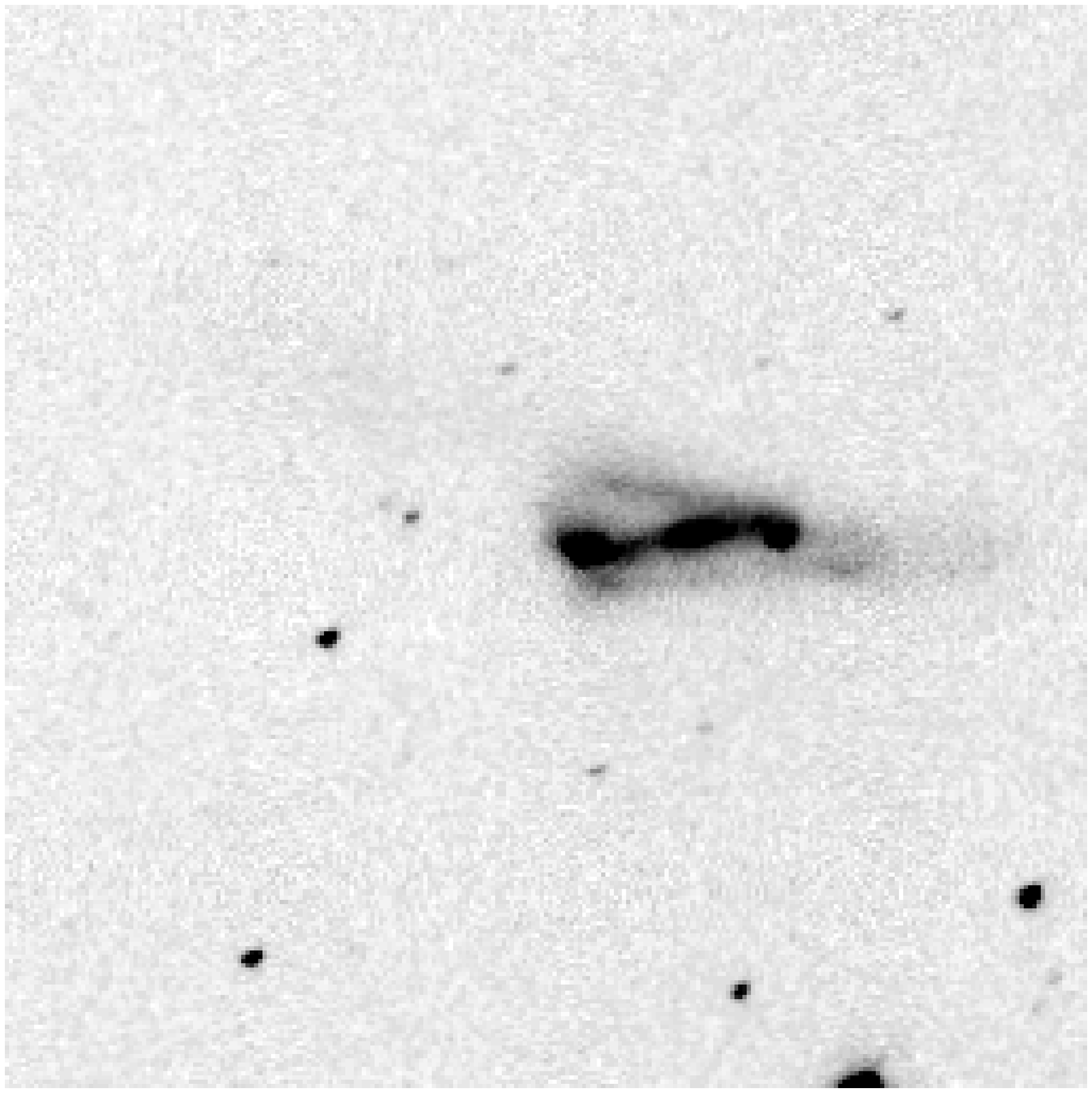}}
\scalebox{0.22}{\includegraphics{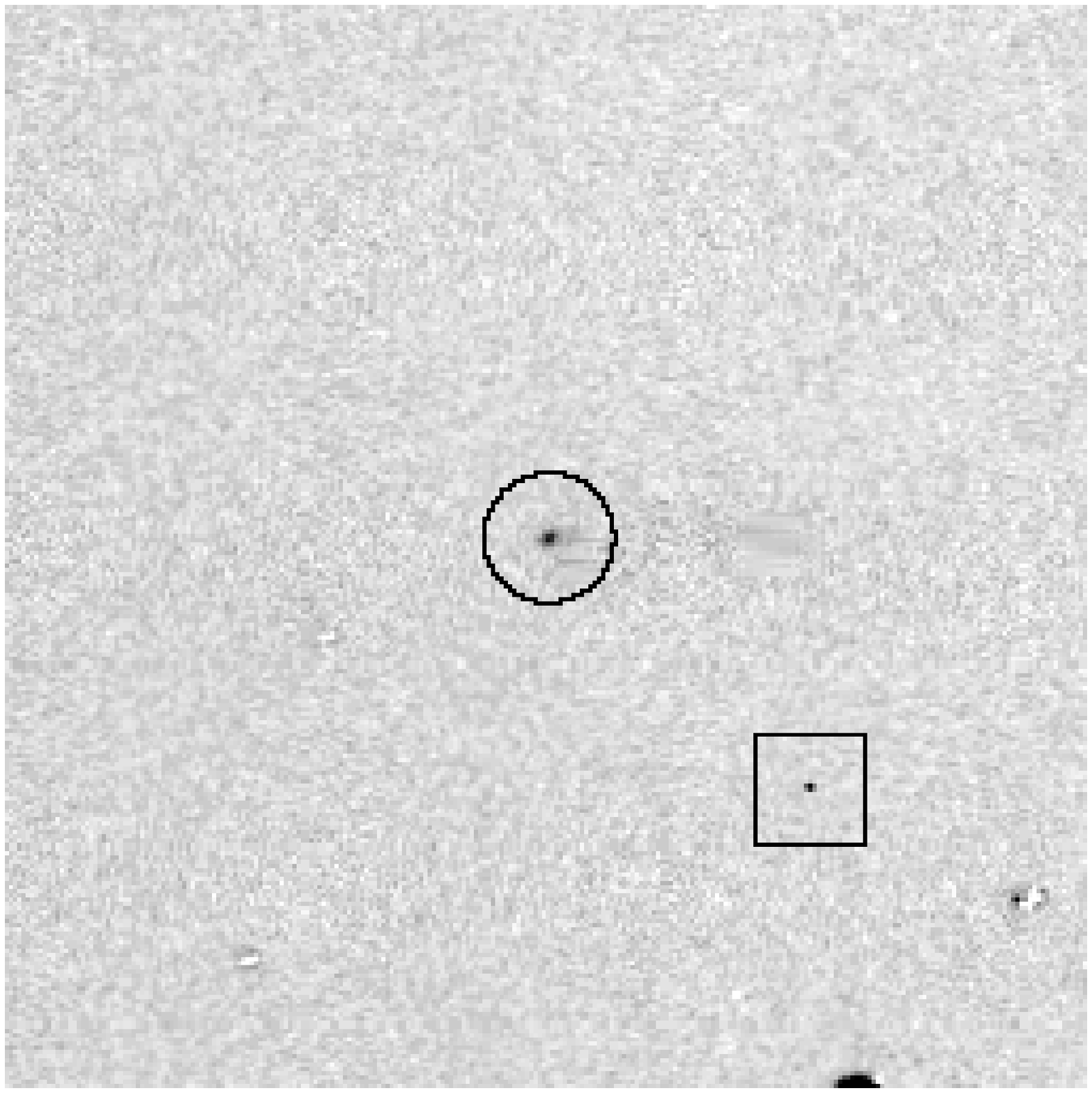}}
\end{center}
\noindent{\bf Fig. 1.} {KAIT images of NGC 523 before {\it (left)} and after
{\it (center)} SN 2001en appeared. The difference image {\it (right)}
shows the SN (circled), along with a cosmic ray (boxed).}
\bigskip

   A Web page describing LOSS is at
http://astro.berkeley.edu/$\sim$bait/kait.html. LOSS found its first supernova
in 1997 --- SN 1997bs, which might not even be a ``genuine'' SN (Van Dyk \etal\
2000). In 1998, mostly during the second half of the year, LOSS discovered 20
SNe, thereby breaking the previous single-year record of 15 held by the Beijing
Astronomical Observatory Supernova Search.  In 1999, LOSS doubled this with 40
SNe. In 2000, LOSS found 38 SNe, even though we spent a significant fraction of
the observing time expanding the database of monitored galaxies rather than
searching for SNe. With this expanded database, LOSS discovered 68 SNe in 2001
and 52 SNe in 2002 through July.  We discovered SN 2000A and SN 2001A, and
hence the first supernova of the new millennium, regardless of one's definition
of the turn of the millennium! During the past few years, KAIT has discovered
{\it about half} of all nearby SNe reported world-wide, from all searches
combined --- and through July 2002 it accounted for well over half (52/86) of
them. Thus, KAIT/LOSS is currently the world's most effective search engine for
nearby SNe.

To further increase the sample of SNe, my group recently decided to coordinate
with Michael Schwartz of the Tenagra Observatory (0.60~m and 0.75~m
telescopes), using largely complementary sets of galaxies (the overlap between
galaxy lists is larger during those months when either observatory often has
bad weather).  Our ``Lick Observatory and Tenagra Observatory Supernova
Search'' (LOTOSS; Schwartz \etal\ 2000) should discover almost all of the nearby
SNe over the accessible areas of the sky. Already, the Tenagra telescopes have
discovered several SNe (e.g., Schwartz \& Li 2001, 2002).  Also, this strategy
leaves KAIT with more time to conduct multicolor follow-up photometry of SNe.

 At Lick and Keck Observatories, we spectroscopically confirm and classify
nearly all of the SNe that other observers haven't already classified. Thus,
the sample suffers from fewer biases than most. The distribution of types
through SN 2002dy is 93 SNe~Ia, 76 SNe~II, 12 SNe~IIn, 1 SN~IIb, 12 SNe~Ib, 16
SNe~Ic, and 6 unknown.  We have started to determine the Hubble types of the
host galaxies of the SNe (van den Bergh, Li, \& Filippenko 2002), as a first
step in the calculation of rates of various types of SNe. Already, our
observations and Monte-Carlo simulations have shown that the rate of
spectroscopically peculiar SNe~Ia is considerably larger than had previously
been thought (Li \etal\ 2001a).

Follow-up observations for the discovered SNe are emphasized during the course
of LOSS. Our goal is to build up a multicolor database for nearby
SNe. Because of the early discoveries of most LOSS SNe, our light curves
usually have good coverage from pre-maximum brightening to post-maximum
decline. Moreover, all LOSS SNe are automatically monitored in
unfiltered mode as a byproduct of our search; these can sometimes be useful for
other studies (e.g., Matheson \etal\ 2001). The positions of SNe in KAIT
early-time images were used to identify the same SNe at very late times in {\it
Hubble Space Telescope} images (Li \etal\ 2002), allowing us to determine
the late-time decline rates.

   LOSS also discovers novae in nearby galaxies (e.g., M31), cataclysmic
variable stars, and occasionally comets (e.g., Li 1998). Although it records
many asteroids, we don't conduct follow-up observations, so most of them are
lost.

\section {SN 2000cx: A Very Weird SN Ia}

High-quality observations of nearby SNe Ia provide valuable information
about their progenitor evolution and the relevant physics.  Analyses of 
samples of well-observed, nearby SNe~Ia enable observers to study the 
differences among SNe~Ia, empirical correlations,  and possible environmental
effects (e.g., Hamuy \etal\ 2000).  It is thus important to expand the sample
of well-observed nearby SNe~Ia. Thus, a substantial fraction of KAIT's time
is devoted to follow-up photometry of bright SNe~Ia.

Moreover, studies of high-redshift SNe~Ia have revealed a surprising
cosmological result, that the expansion of the Universe is currently
accelerating, perhaps due to a nonzero cosmological constant (e.g., Riess \etal\
1998, 2001; Perlmutter \etal\ 1999). This result, however, is based on the
assumption that there are no significant differences between SNe~Ia at high
redshift and their low-redshift counterparts. In particular, we rely on the
luminosity/light-curve correlation, as quantified in a number of ways (e.g.,
Riess \etal\ 1998); Phillips \etal\ 1999; Perlmutter \etal\ 1999), to
``standardize" the luminosities of different SNe~Ia. But what if some SNe~Ia
don't conform with this correlation? If there are more of them at high
redshifts than at low redshifts, systematic errors may creep into the
analysis. We need to find and investigate such objects at low redshifts.

SN 2000cx in the S0 galaxy NGC 524 is a case in point. It was discovered and
confirmed by LOSS in July 2000, and at became the brightest SN of the year
2000. A follow-up program of multicolor photometry and spectroscopy was
established at Lick Observatory; photometry of SN 2000cx was also gathered at
the Wise Observatory in Israel. The results are described in detail by Li
\etal\ (2001b); here I summarize the main points.

   A very peculiar object, SN 200cx is, indeed, unique among all known
SNe~Ia. The light curves cannot be fit well by any of the fitting techniques
currently available (e.g., MLCS and the stretch method); see Figure 2.  There
is an apparent asymmetry in the rising and declining parts of the $B$-band
light curve, while there is a unique ``shoulder''-like evolution in the
$V$-band light curve. The $R$-band and $I$-band light curves have relatively
weak second maxima.  In all $BVRI$ passbands the late-time decline rates are
relatively large compared to other SNe~Ia.

\begin{center}
\rotatebox{270}{
 \scalebox{0.5}{\includegraphics{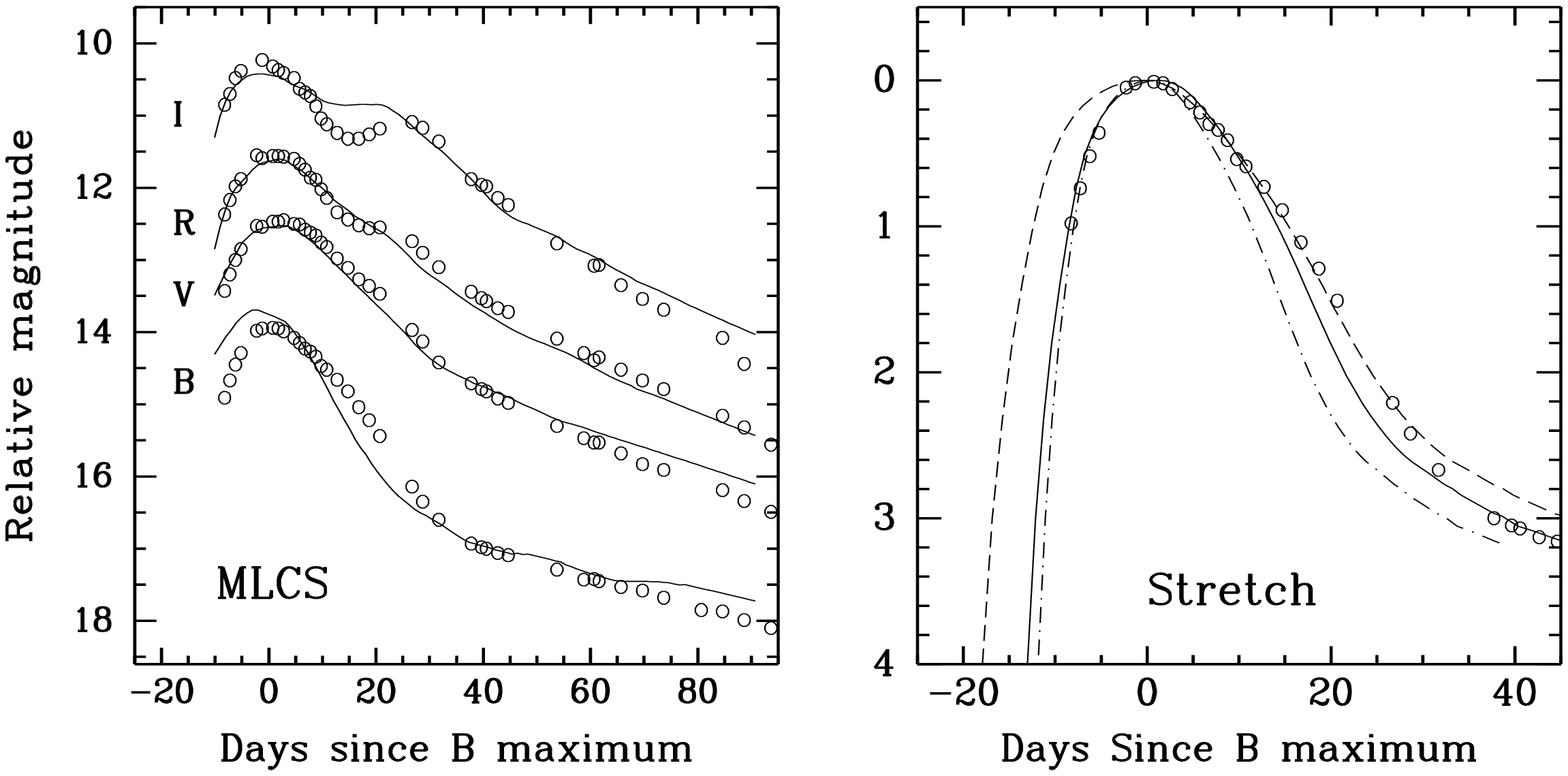}}}
\end{center}
\noindent {\bf Fig. 2.}
{The MLCS fit (Riess \etal\ 1998; {\it left 
panel}) and the stretch method fit
(Perlmutter \etal\ 1999; {\it right panel}) for 
SN 2000cx. The MLCS fit is the worst we had
ever seen through 2000. For the stretch method fit, the solid line is the fit to
all the data points from $t$ = $-$8 to 32 days, the dash-dotted line
uses only the premaximum datapoints,  and the dashed line only the postmaximum
datapoints. The three fits give very different stretch factors. 
From Li \etal\ (2001b).}
\bigskip

 SN 2000cx has the reddest $(B - V)_0$ color before $t \approx$ 7 days among
several SNe~Ia, and it subsequently has a peculiar plateau phase where $(B - V)_0$
remains at 0.3 mag until $t$ = 15 days. The late-time $(B - V)_0$ evolution
of SN 2000cx is found to be rather blue, and is inconsistent with the fit
proposed by Lira (1995) and Phillips \etal\ (1999). SN 2000cx also has very 
blue $(V - R)_0$ and $(V - I)_0$ colors compared with other SNe~Ia.

  Our earliest spectrum of SN 2000cx ($t = -3$ days) reveals remarkable
resemblance to those of SN 1991T-like objects, with prominent Fe~III lines and
weak Si~II lines. As in the case of SN 1991T, Si~II lines strengthened around
the time of maximum brightness. However, the subsequent spectral evolution of
SN 2000cx is quite different from that of SN 1991T.  The Fe~III and Si~II lines
remain strong, and the Fe~II lines remain weak, in the spectra of SN 2000cx
until $t \approx$ 20 days, indicating that the excitation stages of iron-peak
elements change relatively slowly in SN 2000cx compared with other SNe~Ia, and
suggesting that the photosphere of SN 2000cx stays hot for a long time. Both
iron-peak and intermediate-mass elements are found to be moving at very high
velocities in SN 2000cx. The $V_{exp}$ measured from the Si~II $\lambda$6355
line shows a peculiar (nearly constant) evolution.

 We find that the delayed detonation model DD3 (Woosley \& Weaver 1994) 
investigated by Pinto \& Eastman (2001) accounts for the observations
of SN 2000cx rather well. This model suggests that SN 2000cx is similar
to SN 1991T, but with a larger $^{56}$Ni production 
and a higher kinetic energy (i.e., greater expansion velocity for the 
ejecta). We emphasize that because of uncertainties in the current theoretical 
models for SNe~Ia, various views should be considered. For example, 
the big difference between SN 2000cx and SN 1991T in their $V, R, $ and $I$ 
light curves may suggest that they are two very different objects.

\section{SN 2002cx: An Even Weirder SN Ia}

But SN 2000cx is not the end of the story, when it comes to peculiar SNe~Ia. A
more recent, even stranger SN~Ia was SN 2002cx, many of whose properties are
the opposite of those of SN 2000cx!  SN 2002cx was discovered in May 2002 by
Wood-Vasey \etal\ (2002) with the Oschin 1.2-m telescope at Palomar Observatory
in unfiltered images. It host galaxy is CGCG 044-035, at a redshift of $cz =
7184$ km s$^{-1}$ (determined from H~II region emission lines).  An optical
spectrum (Matheson \etal\ 2002) identified the SN as a peculiar SN 1991T-like
event at about a week before maximum brightness, but the object is very {\it
underluminous} (instead of somewhat overluminous) compared with normal
SNe~Ia. The Si~II $\lambda$6355 and Ca~II H \& K lines are extremely weak or
absent, but the Fe~III lines at 4300~\AA\ and 5000~\AA\ are present and
indicate very low expansion velocity --- only half that of normal SNe~Ia.

Recognizing the uniqueness of SN 2002cx shortly after its discovery, we
established a follow-up program of multicolor photometry Lick Observatory.
Spectra of the SN were obtained with the Fred L. Whipple Observatory 1.5-m
telescope and also with the Keck 10-m telescopes. Li \etal\ (2003) discuss the
results in detail; here I provide only a brief summary.

    Besides being subluminous by $\sim 2$ mag at all optical wavelengths
relative to normal SNe~Ia (implying that only a small amount of $^{56}$Ni was
produced), SN 2002cx has peculiar photometric evolution (Fig. 3). In the $B$
band it has a decline rate of $\Delta m_{15}(B) = 1.29\pm0.11$ mag, similar to
those of SN 1994D and SN 1999ac, but it is less luminous by $\sim 1.4$ mag than
SN 1994D and SN 1999ac. The $R$ band has a broad peak, and the $I$ band
has a unique plateau that lasts until about 20 days after $B$ maximum. The
late-time decline is rather slow in all $BVRI$ bands.  The $(B - V)$ color
evolution is nearly normal, but the $(V - R)$ and $(V - I)$ colors are very
red.

\begin{center}
\rotatebox{270}{
\scalebox{0.54}{\includegraphics{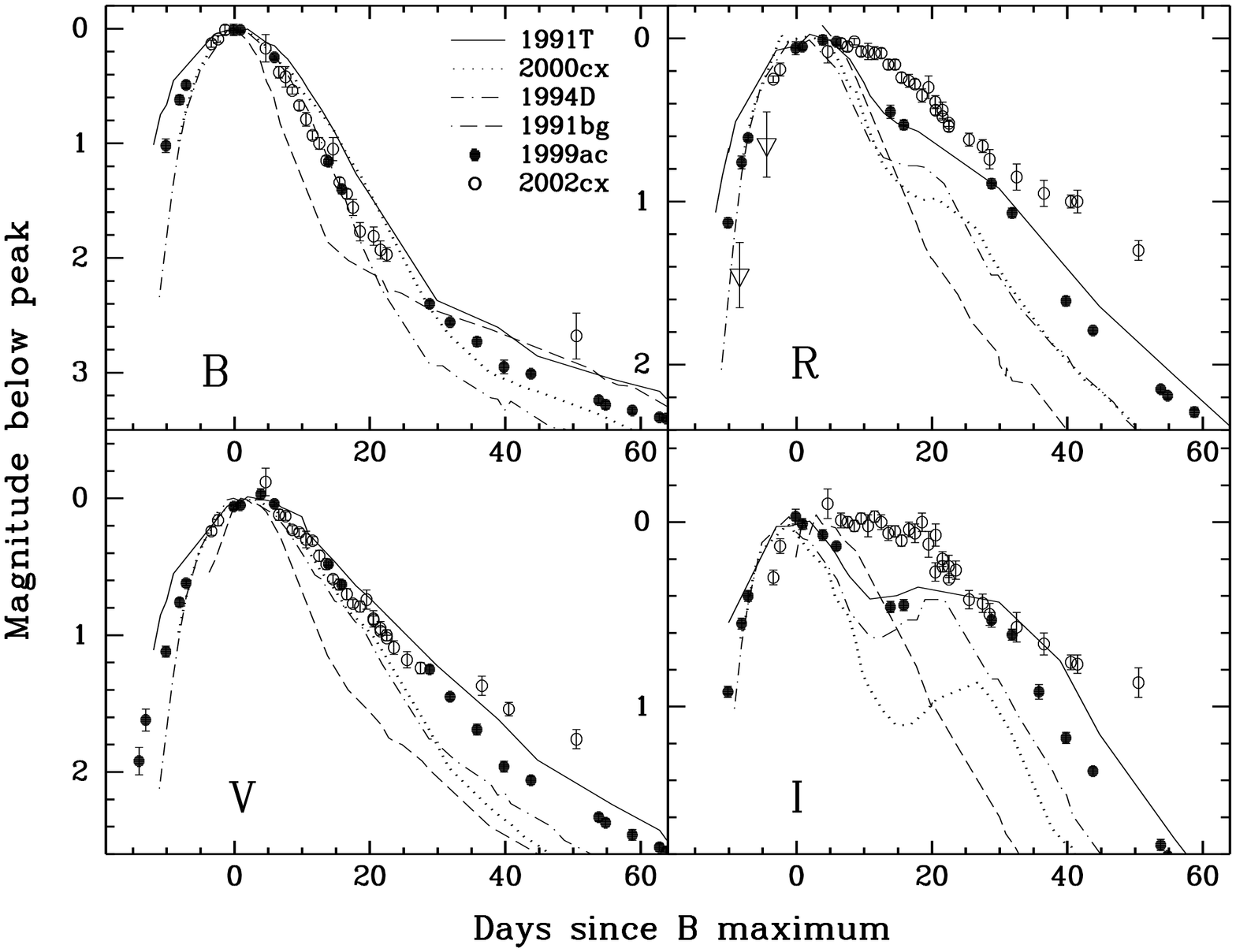}}}
\end{center}
\noindent {\bf Fig. 3.}
{Comparison (Li \etal\ 2003) between the {\it B, V, R, I} light curves 
of SN 2002cx and those of SN 1991T (Lira \etal\ 1998), SN 2000cx (Li \etal\
2001b), SN 1994D (Richmond \etal\ 1995), SN 1991bg (Filippenko \etal\ 1992;
Leibundgut \etal\ 1993), and SN 1999ac (Li \etal, in preparation). 
All light curves are shifted in time and peak magnitude to match those 
of SN 2002cx.}
\bigskip

   The premaximum spectrum of SN 2002cx resembles those of SN 1991T-like
objects, but with extremely low expansion velocities, the lowest ever measured
for a SN~Ia. The spectral evolution is dominated by Fe-group element lines,
with very weak intermediate-mass element features.  The nebular phase was
reached unprecedently soon after maximum, despite the low velocity of the
ejecta, implying that the ejected mass is low. The nebular-phase spectrum is
also quite different from those of other SNe~Ia (Fig. 4); there are mysterious
emission lines near 7000~\AA\ around 3 weeks after maximum brightness, and
other differences as well.  At late times, the spectrum is dominated by very
narrow Fe~II and Co~II lines, and the object is very red.

   SN 2002cx is inconsistent with the observed SN~Ia decline rate
vs. luminosity relation, or the spectral vs. photometric sequence. No existing
theoretical model successfully explains all observed aspects of SN 2002cx,
though the pulsating delayed detonation of a Chandrasekhar-mass white dwarf or
the He detonation of a sub-Chandra white dwarf have some promising
characteristics and should be pursued further.

\begin{center}
\rotatebox{270}{
\scalebox{0.52}{\includegraphics{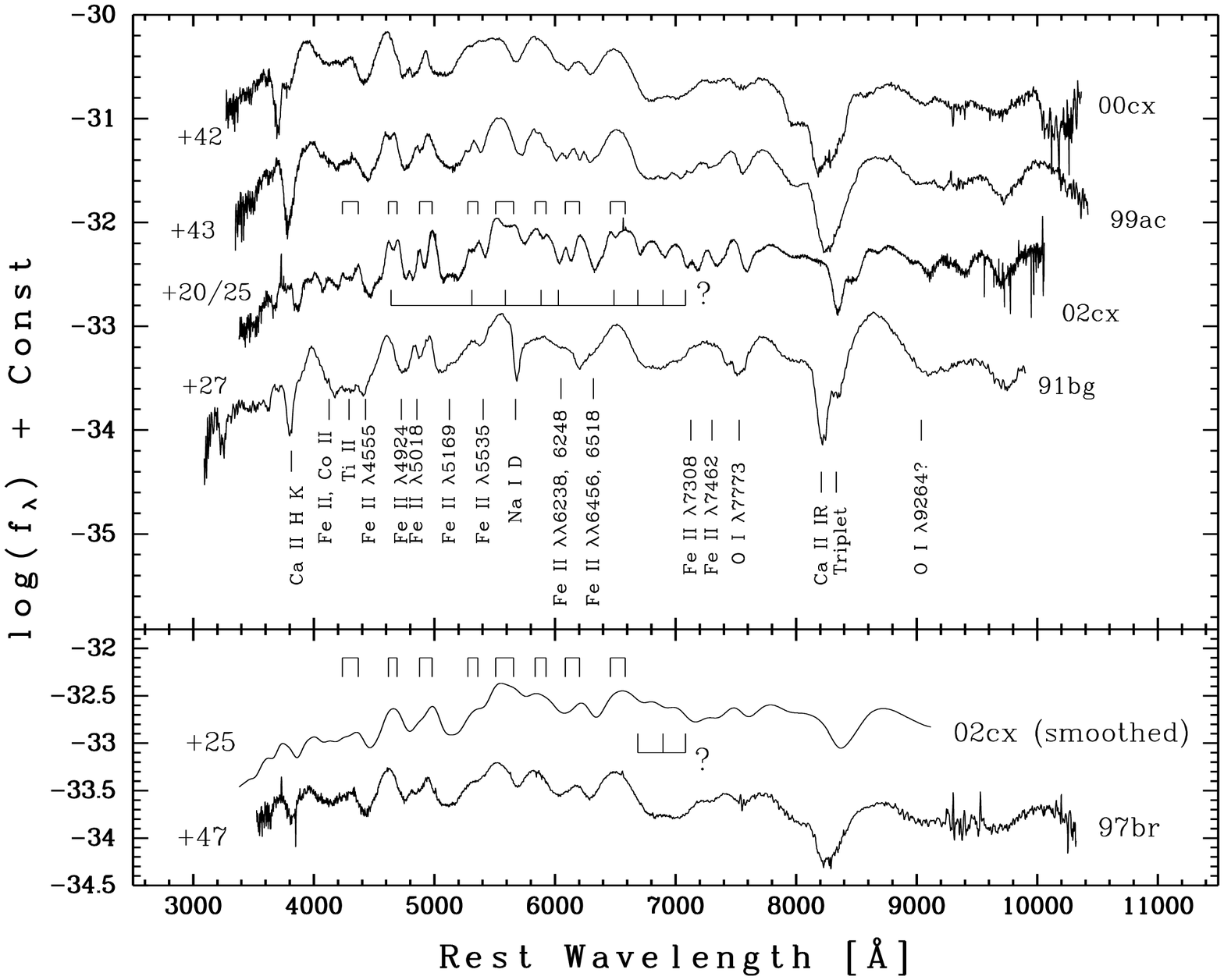}}}
\end{center}
\noindent {\bf Fig. 4.}
{The spectrum of SN 2002cx at $t$ = +20/25~d, shown with spectra of other
SNe~Ia at older ages (Li \etal\ 2003). 
The {\it upper panel} shows the line identifications and the
comparison of the spectra. The pairs of short vertical lines above the SN
2002cx spectrum mark possible ``double peaks," while these below the SN 2002cx
spectrum mark possible additional resolved lines (compared with other SNe~Ia).
The lower panel shows the comparison between the $t$ = +25~d spectrum of SN
2002cx after convolving with a Gaussian function with $\sigma$ = 2,500 km
s$^{-1}$, and the day +47 spectrum of SN 1997br.  Note that although the
``double peaks" are gone, additional features seem to be present around
7000~\AA\ in the spectrum of SN 2002cx.}

\section{Studies of Type II Supernovae}

We have also used KAIT to obtain excellent light curves of SNe~II, with
complementary spectra obtained at Lick Observatory and elsewhere. These are
being used to study the physical properties of SNe~II, and also to derive
distances through the expanding photosphere method (EPM), a variant of the
Baade (1926) method used to measure distances to variable stars.

In two detailed studies, we derived EPM distances to SN 1999em ($D = 8.2 \pm
0.6$ Mpc; Leonard \etal\ 2002a) and SN 1999gi ($D = 11.1^{+2.0}_{-1.8}$;
Leonard \etal\ 2002b).  In addition to its cosmological use, knowing the EPM
distance to SN~1999gi allowed us to set constraints on the upper mass limit of
its progenitor star of $15^{+5}_{-3} {\rm \ M}_{\odot}$, through the analysis
of prediscovery images. This is substantially less restrictive than the upper
mass limit ($9^{+3}_{-2} {\rm \ M}_{\odot}$) recently found in the same manner
by Smartt \etal\ (2001, 2002).  The increased upper limit results mainly from
the larger distance derived through EPM than was assumed by the Smartt \etal\
(2001, 2002) analyses, which relied on less precise (and less recent) distance
measurements to NGC~3184.

  We have also obtained high signal-to-noise ratio spectropolarimetry of some
SNe~II-P with the Keck 10-m and Lick 3-m telescopes (Leonard \etal\ 2001). At
early times, SNe~II-P appear to be polarized very little, suggesting that any
departures from spherical symmetry are small. This is encouraging news for
those who attempt to derive EPM distances for SNe~II-P: Unlike the empirically
based method used to measure distances to SNe~Ia, distances derived to SNe~II-P
rely on the assumption of a spherically symmetric flux distribution during the
early stages of development (i.e., the plateau). We plan to obtain
spectropolarimetry of additional SNe~II-P, in order to much more thoroughly
test the fundamental assumption of spherical symmetry in EPM.

However, multi-epoch spectropolarimetry shows that the polarization increased
with time (Fig. 5a), implying a substantially spherical geometry at early times
that becomes more aspherical at late times when the deepest layers of the
ejecta are revealed. In addition, our data on other core-collapse SNe indicates
large polarizations for objects that have lost much of their envelope prior to
exploding (e.g., SN~Ic 2002ap, Fig. 5b; see below, and Leonard \etal\ 2002c).
For core-collapse events, then, it seems that the closer we probe to the heart
of the explosion, the greater the polarization and, hence, the asymmetry.  The
current speculation is that the presence of a thick hydrogen envelope dampens
the observed asymmetry.

\section {The Peculiar SN Ic 2002ap}

Although core-collapse SNe present a wide range of spectral and
photometric properties, there is growing consensus that much of this variety is
due to the state of the progenitor star's hydrogen and helium envelopes at the
time of explosion. Those stars with massive, intact envelopes produce
Type II-plateau SNe, those that have lost their entire hydrogen envelope (perhaps
through stellar winds or mass transfer to a companion) result in SNe~Ib, and
those that have been stripped of both hydrogen and most (or all) of their
helium produce SNe~Ic; see Filippenko (1997) for a general review.

\begin{center}
\rotatebox{0}{
\scalebox{0.45}{\includegraphics{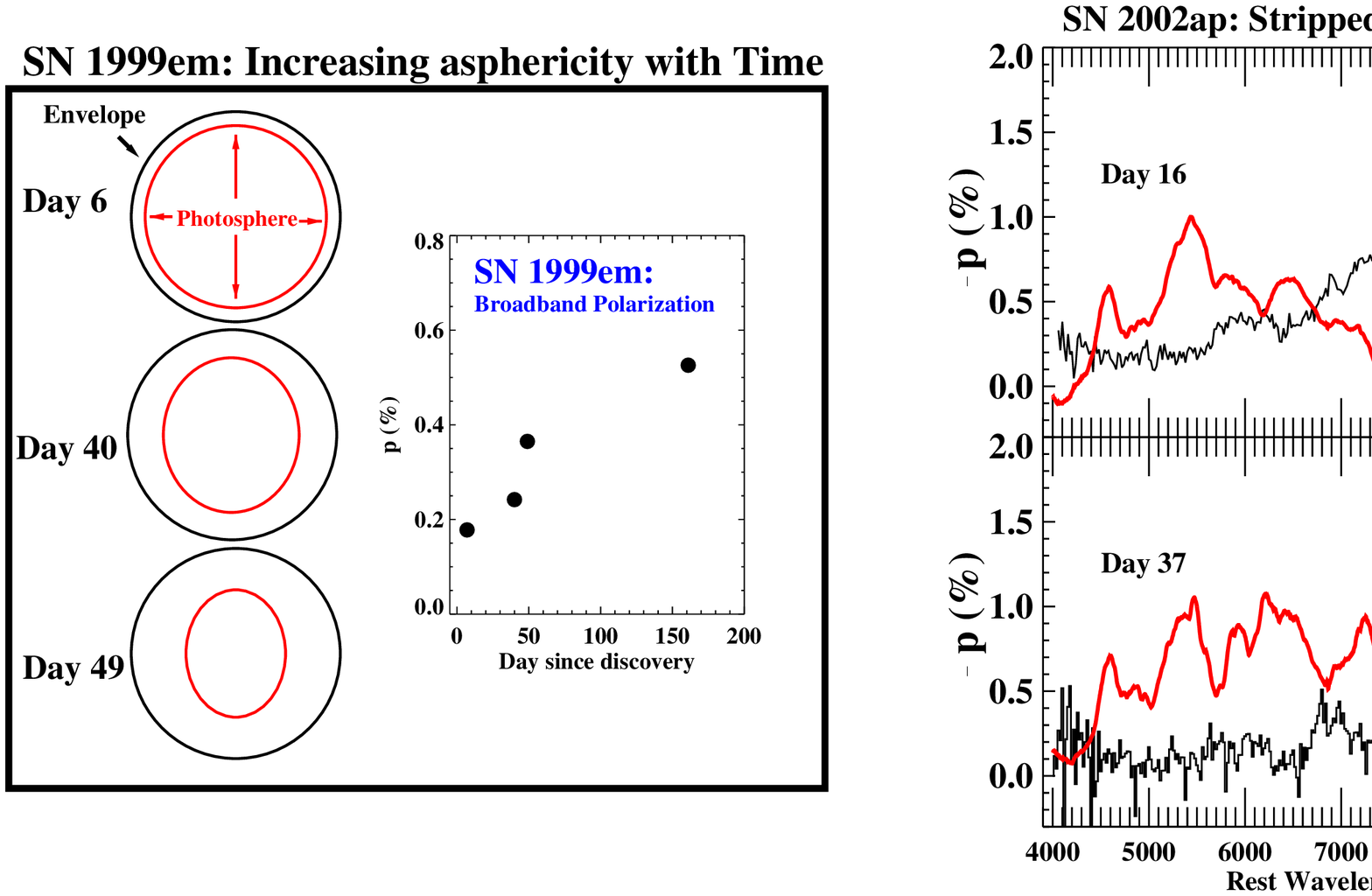}}}
\end{center}
\noindent {\bf Fig. 5.}
{(a) {\it (left)} The temporal increase in the polarization of the Type
II-P SN 1999em suggests greater asphericity deeper into the ejecta.
(b) {\it (right)}
 Polarization level ({\it thin noisy
lines}) of the peculiar SN~Ic 2002ap, with relative flux ({\it thick smooth
lines}) overplotted for comparison of features (Leonard \etal\ 2002c).
}
\bigskip

Recently, a new subclass of objects has emerged whose members generically
resemble SNe~Ic (no hydrogen or obvious helium spectral features), but, unlike
traditional SNe~Ic, have spectra characterized by unusually broad features at
early times, indicating velocities in excess of $\sim 30,000$ km s$^{-1}$.  A
few also possess inferred kinetic energies exceeding that of ``normal''
core-collapse SNe by more than a factor of 10 (see, e.g., Nomoto \etal\ 2001).
These objects are colloquially referred to as ``hypernovae,'' although not all
of them are clearly more luminous or energetic than normal SNe~Ic.

  Intense interest in hypernovae has been sparked not only by their peculiar
spectral features, but also by the strong spatial and temporal association
between the brightest and most energetic of these events, SN 1998bw, and the
$\gamma$-ray burst (GRB) 980425 (e.g., Galama \etal\ 1998).  There are only a
few generally accepted members of this rare class (e.g., SN~1997dq, SN~1997ef).
A related subclass of SNe exhibits many of the characteristics of these
objects, but with hydrogen present in the spectra; the clearest examples are SN
1997cy and SN 1999E (Germany \etal\ 2000; Turatto \etal\ 2000; Filippenko 2000,
and references therein), and they, too, are sometimes called hypernovae. The
hydrogen emission probably comes from the interaction of relatively
hydrogen-poor ejecta with circumstellar gas previously expelled by the
progenitor star (and richer in hydrogen than the remaining parts of the
progenitor).

  SN~2002ap in M74 (Mazzali \etal\ 2002) is one of the most recent examples of a
peculiar SN~Ic of this kind. So far, it has been the brightest supernova of the
year 2002. Though not discovered by LOSS (it went off between two KAIT
observations, but was discovered during that interval by Yoji Hirose [IAUC
7810]), KAIT's observations set a useful limit on the explosion date.

Figure 6 shows the spectrum of SN~2002ap obtained after the SN reappeared
following solar conjunction, about 5 months after maximum brightness. It
is characterized by strong emission lines of intermediate-mass
elements superimposed on a weak continuum, implying that the SN has entered the
nebular phase. Unusual narrow lines are visible on top of some of the
broad-line profiles, including especially those of [O~I] $\lambda\lambda$6300,
6364 and Mg~I] $\lambda$4571.

\begin{center}
\rotatebox{90}{
\scalebox{0.58}{\includegraphics{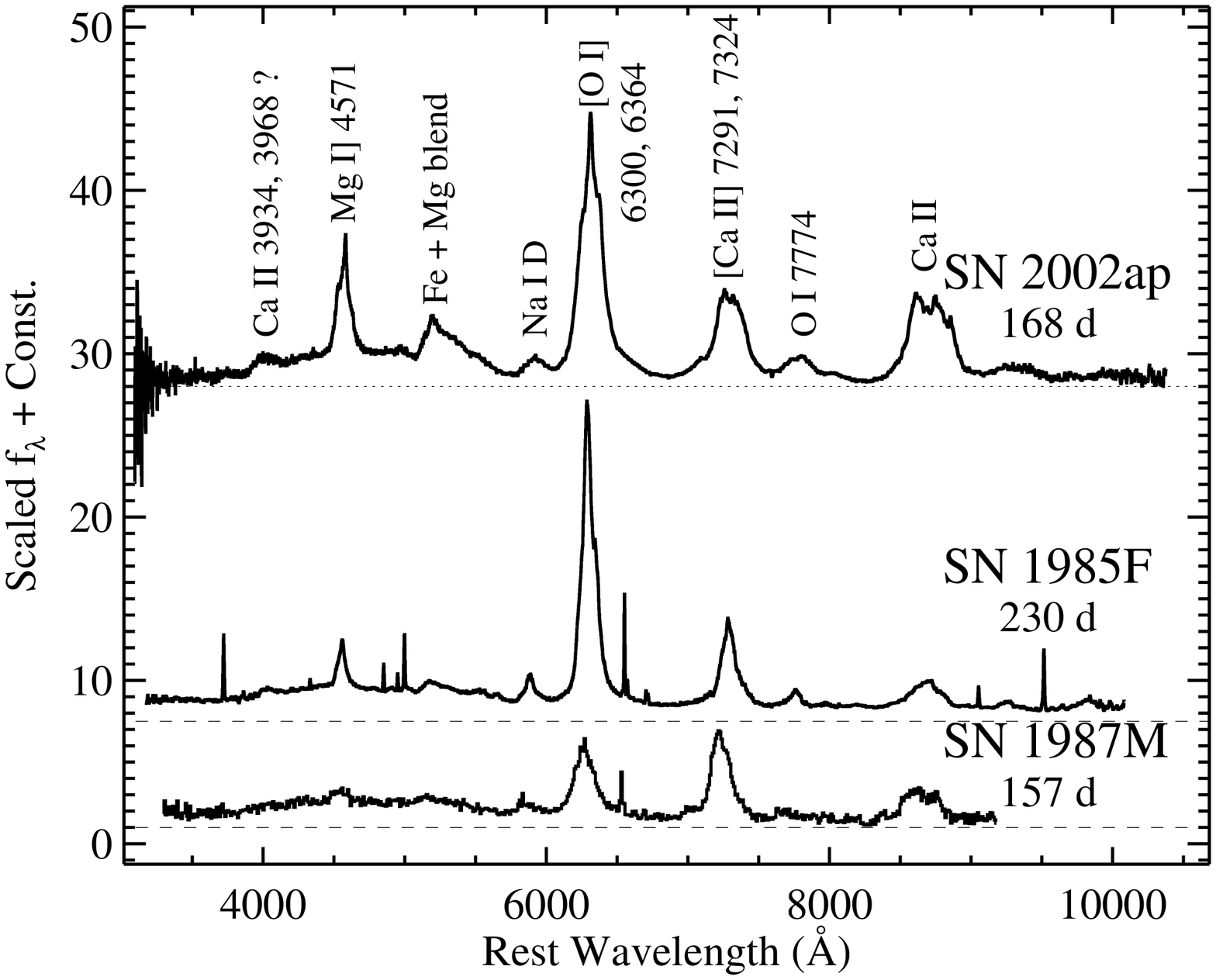}}}
\end{center}
\noindent {\bf Fig. 6.}
{SN~2002ap ({\it top}) in the nebular phase, compared with the SN~IIb/Ib/Ic
1985F ({\it middle}) and the ``ordinary''
Type Ic SN~1987M ({\it bottom}) at similar epochs; the estimated day since
explosion is indicated. The exact spectral classification of SN 1985F is unknown,
since it was discovered long after maximum brightness (see Filippenko \&
Sargent 1986; Filippenko 1997), but it provides the closest match we could find
to the spectrum of SN 2002ap. The spectra are scaled so that the height of the
[Ca~II] $\lambda\lambda$7319, 7324 blend is approximately the same in
all three cases.}
\bigskip

Although there is no guarantee that the three objects were at the same stage of
their development (they may evolve at different physical rates, even if they
are approximately the same calendar age), one can see in Figure 6 that relative
to SN 1987M, a typical SN~Ic, SN 2002ap has {\it much} stronger [O~I] and Mg~I]
emission. The only SN~Ib/Ic we have found comparable to SN 2002ap is SN 1985F,
as shown. However, SN 2002ap exhibits a larger Mg~I] $\lambda$4571 to [O~I]
$\lambda\lambda$6300, 6364 ratio than that of SN 1985F, suggesting that we are
seeing even closer to the O-Ne-Mg layer in SN 2002ap. Qualitatively, this
supports the hypothesis that SN 2002ap (and perhaps other peculiar examples of
the SN~Ic subclass) have progenitors that are even more highly stripped than
normal SNe~Ic. There might be other factors to consider as well, but the
sequence II-P $\rightarrow$ IIb $\rightarrow$ Ib $\rightarrow$ Ib/c
$\rightarrow$ Ic $\rightarrow$ Ic-pec may fundamentally be one dominated by the
degree to which the envelope of the progenitor has been stripped.

\section {The Progenitors of Core-Collapse SNe}

Identifying the massive progenitor stars that give rise to core-collapse
SNe is one of the main pursuits of supernova and stellar evolution
studies.  Using ground-based images of recent, nearby SNe obtained primarily
with KAIT, astrometry from the Two Micron All Sky Survey, and archival images
from the {\sl Hubble Space Telescope}, we have attempted the direct
identification of the progenitors of 16 Type II and Type Ib/c SNe (Van Dyk
\etal\ 2002).  

We may have identified the progenitors of the Type II SNe 1999br
in NGC 4900, 1999ev in NGC 4274, and 2001du in NGC 1365 as supergiant stars
with $M^0_V\approx -6$ mag in all three cases.  We may have also identified the
progenitors of the Type Ib SNe 2001B in IC 391 and 2001is in NGC 1961 as very
luminous supergiants with $M^0_V \approx -8$ to $-9$ mag, and possibly the
progenitor of the Type Ic SN 1999bu in NGC 3786 as a supergiant with
$M^0_V\approx -7.5$ mag.  

Additionally, we have recovered at late times SNe
1999dn in NGC 7714, 2000C in NGC 2415, and 2000ew in NGC 3810, although none of
these had detectable progenitors on pre-supernova images.  In fact, for the
remaining SNe only limits can be placed on the absolute magnitude and color
(when available) of the progenitor.  The detected Type II progenitors and
limits are consistent with red supergiants as progenitor stars, although
possibly not as red as we had expected.  Our results for the SNe~Ib/c do
not strongly constrain either Wolf-Rayet stars or massive interacting binary
systems as progenitors. 

\section {Acknowledgments}

   I am grateful to the Committee on Research (U.C. Berkeley) and the
conference organizers for providing partial travel support to attend this
meeting.  My recent research on SNe has been financed by the US National
Science Foundation, most recently through grant AST-9987438, as well as by NASA
grants AR-8754, GO-9114, GO-9428, and AR-9529 from the Space Telescope Science
Institute, which is operated by AURA, Inc., under NASA Contract NAS5-26555.
KAIT and its associated science have been made possible with funding or
donations from NSF, NASA, the Sylvia and Jim Katzman Foundation, Sun
Microsystems Inc., Lick Observatory, the Hewlett-Packard Company, Photometrics
Ltd., AutoScope Corporation, and the University of California. I thank
R. Foley, D. C. Leonard, and W. D. Li for assistance with the figures.

\end{document}